\documentclass[12pt,preprint]{aastex}
\usepackage{emulateapj5}
\usepackage{onecolfloat}
\usepackage{apjfonts}
\usepackage{subfigure}  

\newcommand{\mincir}{\raise-3.truept\hbox{\rlap{\hbox{\hspace{3.truept}
$\sim$}}\raise4.truept\hbox{\hspace{3.truept}$<$}\ }}
\newcommand{\magcir}{\raise-3.truept\hbox{\rlap{\hbox{\hspace{3.truept}
$\sim$}}\raise4.truept\hbox{\hspace{3.truept}$>$}\ }}
\newcommand{\minmag}{\raise-2.truept\hbox{\rlap{\hbox{$<$}}
\raise6.truept\hbox{$>$}\ }}
\newcommand{\hm}{\,h^{-1}{\rm Mpc}}
\newcommand{\be}{\begin{equation}} 
\newcommand{\ee}{\end{equation}} 
\newcommand{\chandra}{\emph{Chandra}} 
\newcommand{\xmm}{XMM-\emph{Newton}} 
 
\begin{document} 
\twocolumn[%
 
\title{ Mismatch between X-ray and emission-weighted temperatures in galaxy
clusters: cosmological implications}
 
\author{ 
  E. Rasia\altaffilmark{1}, 
  P. Mazzotta \altaffilmark{2,3}, 
  S. Borgani\altaffilmark{4,5}, 
  L. Moscardini\altaffilmark{6},  
  K. Dolag\altaffilmark{1,6},   
  G. Tormen\altaffilmark{1},  
  A. Diaferio\altaffilmark{7}, 
  G. Murante\altaffilmark{8}
}  
 
\begin{abstract} 
The thermal properties of hydrodynamical simulations of galaxy
clusters are usually compared to observations by relying on the
emission-weighted temperature $T_{\rm ew}$, instead of on the
spectroscopic X-ray temperature $T_{\rm spec}$, which is obtained by
actual observational data. In a recent paper Mazzotta et al. show
that, if the cluster is thermally complex, $T_{\rm ew}$ fails at
reproducing $T_{\rm spec}$, and propose a new formula, the
spectroscopic-like temperature, $T_{\rm sl}$, which approximates
$T_{\rm spec}$ better than a few per cent. By analyzing a set of
hydrodynamical simulations of galaxy clusters, we find that $T_{\rm
sl}$ is lower than $T_{\rm ew}$ by 20--30 per cent.  As a consequence,
the normalization of the $M$--$T_{\rm sl}$ relation from the
simulations is larger than the observed one by about 50 per cent. If
masses in simulated clusters are estimated by following the same
assumptions of hydrostatic equilibrium and $\beta$--model gas density
profile, as often done for observed clusters, then the $M$--$T$
relation decreases by about 40 per cent, and significantly reduces its
scatter. Based on this result, we conclude that using the observed
$M$--$T$ relation to infer the amplitude of the power spectrum from
the X--ray temperature function could bias low $\sigma_8$ by 10-20 per
cent. This may alleviate the tension between the value of $\sigma_8$
inferred from the cluster number density and those from cosmic
microwave background and large scale structure.
\end{abstract} 
 

]
    
\altaffiltext{1}{Dipartimento di Astronomia, 
Universit\`a di Padova, vicolo dell'Osservatorio 2, I-35122 Padova,
Italy (rasia,kdolag,tormen@pd.astro.it)}

\altaffiltext{2}{Dipartimento di Fisica, Universit\`a di Roma Tor Vergata,  
via della Ricerca Scientifica 1, I-00133 Roma, Italy
(pasquale.mazzotta@roma2.infn.it)}

\altaffiltext{3}{Harvard-Smithsonian Center for Astrophysics, 60 Garden  
Street, Cambridge, MA02138, USA}

\altaffiltext{4}{Dipartimento di Astronomia dell'Universit\`a di Trieste,  
via Tiepolo 11, I-34131 Trieste, Italy (borgani@ts.astro.it)}

\altaffiltext{5} {INFN -- National Institute for 
Nuclear Physics, Trieste, Italy}

\altaffiltext{6}{Dipartimento di Astronomia, Universit\`a di Bologna,  
via Ranzani 1, I-40127 Bologna, Italy (lauro.moscardini@unibo.it)}

\altaffiltext{7}{ Dipartimento di Fisica Generale ``Amedeo Avogadro'', 
Universit\`a di Torino, via Giuria 1, I-10125, Torino, Italy
(diaferio@ph.unito.it) }

\altaffiltext{8}{INAF, Osservatorio Astronomico di Pino Torinese,  
Strada Osservatorio 20, I-10025 Pino Torinese, Italy
(murante@to.astro.it)}
 
\section{Introduction}\label{sec1} 

Thanks to their position at the top of the hierarchy of cosmic structure
formation, galaxy clusters represent powerful tools to constrain cosmological
parameters, like the present value of the matter density parameter
$\Omega_{0m}$, and the normalization of the primordial power spectrum, usually
expressed in terms of $\sigma_8$, the r.m.s. matter fluctuation in a sphere of
radius of $8 h^{-1} $ Mpc (Rosati et al. 2002; Voit 2004 for recent
reviews). The fundamental quantity entering theoretical models is the cluster
mass, but from the observational point of view its high-precision measurement
is still precluded by the presence of different systematic effects affecting
the lensing data and the dynamical models.  A possible short cut to overcome
this difficulty is represented by the determination of the cluster temperature
from X-ray observations. In fact simple theoretical arguments, supported by
numerical hydro-N-body simulations, suggest the existence for virialized
gravitational systems of a tight relation between mass $M$ and temperature
$T$.
 
The enormous progresses in the angular and spectral resolutions of the latest
generation X-ray satellites, \chandra ~ and XMM-Newton, allowed us to obtain
temperature measurements for extended sets of clusters, showing in many cases
the presence of a rather complex thermal structure.  The theoretical
interpretation of these high-quality data requires state-of-the art numerical
simulations, in which the physical processes acting on the gas component
(radiative cooling, non-gravitational heating, etc.) are faithfully
reproduced.  The comparison between observations and simulations is, however,
not always simple. As for X-ray data, the cluster gas temperature is obtained
by fitting the source spectrum to a suitable thermal model. This procedure
cannot be reproduced in the analysis of simulations, without using devoted
software packages (like X-MAS; Gardini et al. 2004), which simulate the
spectral properties of the simulated clusters taking into account the
instrument response, as well as sky and instrumental backgrounds. To overcome
this problem, it has been of common use to resort to the emission-weighted
temperature, \be T_{\rm ew}\equiv \frac{\int \Lambda(T) n^2 T dV} {\int
\Lambda(T) n^2 dV}\,.
\label{eq:tew} 
\ee  
Here $n$ is the gas density and $\Lambda(T)$ is the cooling function,
for which most of the works in the literature use $\Lambda(T)\propto
\sqrt {T}$, assuming bremsstrahlung emission. 
  
Recently, Mazzotta et al. (2004, M04 hereafter) demonstrated that, if
clusters have a complex thermal structure, $T_{\rm ew}$ always
overestimates the spectroscopic temperature $T_{\rm spec}$ obtained
from X-ray observations. The discrepancy between $T_{\rm ew}$ and
$T_{\rm spec}$ strongly depends on the complexity of the cluster
thermal structure (see also Mathiesen \& Evrard 2001; Gardini et
al. 2004).  M04 proposed a new definition of temperature to be
implemented in the analysis of simulations, the spectroscopic-like
temperature,
\be  
T_{\rm sl}\equiv \frac{\int n^2T^{a}/T^{1/2} dV }  
{\int n^2T^{a}/T^{3/2} dV}\,. 
\label{eq:tsl}  
\ee 
When applied to simulated spectra of clusters hotter than 2-3 keV observed by
\chandra ~ or \xmm ~, this equation, with $a=0.75$, yields a temperature
within a few per cent from $T_{\rm spec}$. 
M04 also showed that this result is almost independent of
the details of the detector response function, as long as its spectral
properties are similar to or worse than the \chandra ~ ones.  In this
way, Eq.~\ref{eq:tsl} can be reliably applied to the detectors on
board of Beppo-SAX and ASCA.

In this Letter we address the implications of using $T_{\rm sl}$ on
the mass--temperature relation and on the X--ray temperature function
(XTF) for an extended set of simulated clusters. These results are
then used to discuss the effect on the estimate of $\sigma_8$ from the
XTF.

\section{The simulated clusters}\label{sec2} 
 
The sample of simulated galaxy clusters used in this paper has been
extracted by the large-scale cosmological hydro-N-body simulation of a
``concordance'' $\Lambda$CDM model ($\Omega_{0m}=0.3$,
$\Omega_{0\Lambda}=0.7$, $\Omega_{0\rm b}=0.019\,h^{-2}$, $h=0.7$,
$\sigma_8=0.8$), presented in Borgani et al. (2004, B04
hereafter). Here we give only a short summary of its characteristics,
and we refer to that paper for more details.  The run, made by using
the massively parallel Tree+SPH code {\small P-GADGET2} (Springel et
al. 2001; Springel 2004, in preparation), follows the evolution of
$480^3$ dark matter particles and an equal number of gas particles in
a periodic cube of size $192 h^{-1}$ Mpc. The mass of the gas
particles is $m_{\rm gas}=6.89 \times 10^8 h^{-1} M_\odot$, while the
the Plummer-equivalent force softening is $7.5 h^{-1}$ kpc at $z=0$.
Besides gravity and hydrodynamics, the simulation includes the
treatment of radiative cooling, the effect of a uniform
time--dependent UV background, a sub--resolution model for star
formation from a multiphase interstellar medium, as well as galactic
winds powered by SN explosions (Springel \& Hernquist 2003).  At $z=0$
we extract a set of 95 temperature-selected clusters (hereafter
Set\#1) with $T_{\rm ew}> 2$ keV, value below which the single
spectroscopic temperature for thermally complex systems cannot be
defined (see discussion in M04).
 
Due to the finite box--size, the largest cluster found in the
cosmological simulation has $T_{\rm ew}\approx 7$ keV.  In order to
extend our analysis to more massive and hotter systems, as required
for the study of the $M$--$T$ relation, we further include 4 more
galaxy clusters (hereafter Set\#2) having $M_{\rm vir}>10^{15} h^{-1}
M_\odot$ and belonging to a different set of hydro-N-body simulations
(Dolag et al. 2004 in preparation). Since these objects have been
obtained by re-simulating at high resolution a patch of a pre-existing
cosmological simulation, they have a better mass resolution, with
$m_{\rm gas}= 1.69 \times 10^{8} h^{-1}M_\odot$. These simulations
have been performed by using the same code with the same choice of the
parameters defining star--formation and feedback.
 
Therefore, our total sample (Set\#1 and \#2) comprises 99 objects
spanning the temperature range $2\mincir T_{\rm ew}\mincir 13$ keV.
 
\section{Results}\label{sec3} 
 
For each cluster in our sample we calculate both the
emission--weighted and the spectroscopic--like temperature.  As for
$T_{\rm ew}$, it is computed as in B04 by applying Eq.~\ref{eq:tew},
with the cooling function estimated in the [0.5--10] keV energy band
so as to account for the typical energy coverage of X--ray detectors.
For similar reasons, we apply Eq.~\ref{eq:tsl} to compute $T_{\rm sl}$
by removing the gas particles with $T<0.5$~keV.  We note that, since
the mean temperature of each cluster here considered is higher than 2
keV, the excluded particles usually provide a small fraction of the
total emission measure. Therefore, this procedure does not
significantly affect the values obtained for both $T_{\rm ew}$ and
$T_{\rm sl}$.
 
\begin{figure}[t]  
\begin{center}  
\includegraphics[height=0.3\textheight,width=0.4\textwidth] 
{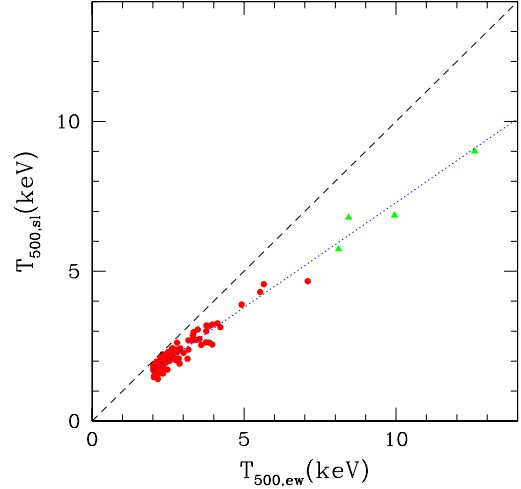}  
\caption{The comparison between emission--weighted temperature, 
$T_{\rm ew}$, and spectroscopic-like temperature, $T_{\rm
sl}$. Circles and triangles refer to objects from Set\#1 and Set\#2,
respectively. The dotted line is the best--fit linear regression (see
Eq.~\ref{eq:TvsT}), while the dashed line corresponds to $T_{\rm
sl}=T_{\rm ew}$.}
\label{fig:TvsT}  
\end{center} 
\end{figure}

In Fig.~\ref{fig:TvsT} we plot the value of $T_{\rm sl}$ versus
$T_{\rm ew}$, which confirms the result of M04, that for all clusters
$T_{\rm sl}$ is smaller than $T_{\rm ew}$.  The distribution of the
points in Fig.~\ref{fig:TvsT} suggests that the two estimators can be
roughly related by a linear relation.
By applying a $\chi^2$ fit, we find the relation,
\be  
T_{\rm sl}= (0.70\pm 0.01) T_{\rm ew} + (0.29\pm 0.05) \,, 
\label{eq:TvsT} 
\ee 
shown as dotted line in Fig.~\ref{fig:TvsT}.  Therefore, for 
objects with temperatures of about 6 keV, the disagreement can be as large 
as 2 keV.  As also discussed by M04, the relative mismatch between $T_{\rm ew}$
and $T_{\rm sl}$ depends only on the thermal structure of the ICM and,
in particular, increases as a function of its complexity.  For this
reason, we expect Eq.~\ref{eq:TvsT} to hold for the physics considered
in our simulation sets. Any physical process (e.g. stronger feedback
or thermal conduction), which produces a more isothermal ICM, is
expected to provide a smaller difference between the two definitions
of temperature.
 
\begin{figure}[t]  
\begin{center}  
\includegraphics[height=0.2\textheight,width=0.48\textwidth] 
{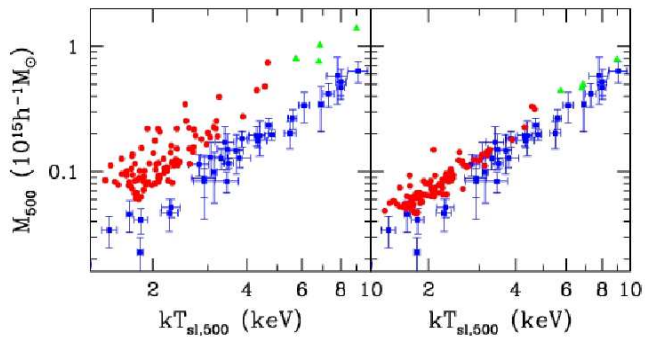} 
\caption{The mass--temperature relation at  
  $\bar\rho/\rho_{\rm cr}=500$, in 
  simulations (filled circles and triangles for Set\#1 and Set\#2, 
  respectively) and for the observational data by Finoguenov et 
  al. (2001, squares with errorbars). The left panel is for the true 
  masses of simulated clusters; the right panel is for masses of 
  simulated clusters estimated by adopting the same procedure applied 
  by Finoguenov et al. to observational data (see text).} 
\label{fig:MvsT}  
\end{center} 
\end{figure}

The existence of a systematic trend between the two different
estimates of the cluster temperature has interesting consequences on
the scaling relation between mass and temperature.  Adiabatic
hydrodynamical simulations (e.g. Evrard et al. 1996; Bryan \& Norman
1998) find a normalization which is about 40 per cent higher than
observed (e.g. Horner et al. 1999; Nevalainen et al. 2001; Finoguenov
et al. 2001; Allen et al. 2001). This result has suggested the
necessity of including in numerical simulations the treatment of extra
physical processes for the gas component, like non--gravitational
heating and radiative cooling. For instance, B04 showed that
considering both effects can be efficient in reducing the
normalization of the simulated $M$--$T$ relation, which however turned
out to be still 20 per cent higher than observed. In that paper, as in
most numerical works on the X-ray properties of galaxy clusters, the
emission--weighted temperature is used to construct the $M$--$T$
relation.  However, since $T_{\rm ew}$ tends to overestimate $T_{\rm
sl}$ by a fair amount, we expect the actual $M$-$T_{\rm sl}$ relation
from simulations to be even more discrepant with respect to
observations.  This discrepancy is indeed apparent in the left panel
of Figure \ref{fig:MvsT}, where the simulated $M$--$T$ relation at
overdensity $\bar\rho/\rho_{\rm cr}=500$\footnote[1]{Here $\rho_{\rm
cr}$ is the cosmological critical density.} is compared to the
observational results by Finoguenov et al. (2001), who measured
temperatures from ASCA data. By modelling the scaling relation with 
\be 
M_{500}={M_0} \left( {T_{500} \over {\rm keV}} \right)^\alpha \,, 
\label{eq:MvsT} 
\ee  
a log-log least--square fitting for the simulated clusters gives
$\alpha =1.66\pm 0.09$ and $\log(M_0/h^{-1}M_\odot)=13.54\pm 0.03$,
while
\newline 
$\log (M_0/h^{-1}M_\odot)=13.60\pm 0.07$ by fixing the slope to the
value $\alpha=1.5$ expected from the condition of hydrostatic
equilibrium.
 
While masses of simulated clusters have been computed here by summing over all
the particles, masses of X--ray observed clusters are usually estimated from
data (e.g., Markevitch 1998; Nevalainen et al. 2000; Finoguenov et al. 2001)
by applying the condition of hydrostatic equilibrium to a spherical gas
distribution described by a $\beta$--model (Cavaliere \& Fusco--Femiano 1976),
with the equation of state having the polytropic form. In this way, the total
self--gravitating mass within the radius $r$ is 
\be 
M(<r)\,=\,1.11\times10^{14} \beta_{\rm fit}\gamma {T(r)\over {\rm keV}} {r\over \hm} {x^2\over 1+x^2}\, h^{-1}M_{\odot} \,. 
\label{eq:hyeq} 
\ee  
Here $T(r)$ is the temperature at $r$, $\beta_{\rm fit}$ is
the fitted slope of the gas density profile, $x=r/r_c$ is the
radial coordinate in units of the core radius $r_c$, and $\gamma$ is
the effective polytropic index.
 
A number of analyses of hydrodynamical simulations of clusters (e.g.,
Muanwong et al. 2002; B04; Rasia et al. 2004; Kay et al. 2004;
cf. also Ascasibar et al. 2003) have shown that Eq.\ref{eq:hyeq}
underestimates the actual cluster mass by about 20 per cent.  Here we
repeat the same analysis as in B04, but using $T_{\rm sl}$ in
Eq.\ref{eq:hyeq} instead of $T_{\rm ew}$.  Although the resulting
$M$--$T$ relation (shown in the right panel of Figure \ref{fig:MvsT})
is still slightly higher than the observed one, we find that the
amplitude of the $M$--$T$ relation decreases quite substantially, with
masses derived from Eq.\ref{eq:hyeq} being on average 40 per cent
smaller than the true ones.  The discrepancy between true and
recovered cluster masses is larger than in B04 as a consequence of
using $T_{\rm sl}$, instead of $T_{\rm ew}$. A comparable bias in the
mass estimates has been also found by Bartelmann \& Steinmetz (1996),
who used a spectroscopic definition of temperatures. Fitting the
simulation results to Eq.\ref{eq:MvsT} gives $\alpha=1.53\pm 0.05$ and
$\log (M_0/h^{-1}M_\odot)=13.40\pm 0.02$, while $\log
(M_0/h^{-1}M_\odot)= 13.41 \pm 0.13$ when fixing $\alpha=1.5$ (we note
that the change of $M_0$ is larger than the 40 per cent average
decrease of the temperature, due to the non--Gaussian distribution of
the scatter in the left panel of Fig.\ref{fig:MvsT}).
 
As already mentioned, the $M$--$T$ relation is one of the key
ingredients in the recipes to extract cosmological parameters from the
cluster X--ray luminosity (XLF) and temperature functions (XTF): for a
fixed value of $\Omega_{0m}$, a larger $M_0$ implies a larger mass for
a fixed temperature and, therefore, a higher normalization of the
power spectrum (e.g., Borgani et al. 2001; Seljak 2002; Pierpaoli et
al. 2003; Henry 2004).
Huterer \& White (2002) suggested an approximation for the
scaling of $\sigma_8$ with $M_0$, involving the matter density
parameter: 
$\Omega_{0m}^{0.6}\sigma_8\propto M_0^{0.53}$ .  
Based on
this relation, we expect that a bias in the mass estimate as large as
that found for the simulated clusters turns into an underestimate of
$\sigma_8$ by about 20 per cent.
 
Besides decreasing $M_0$, estimating cluster masses from
Eq.\ref{eq:hyeq} also significantly reduces the scatter in the
$M$--$T$ relation from about 30 per cent to about 16 per cent, as
evident from Figure \ref{fig:MvsT}. This witnesses that the complex
structure of the ICM enters in determining the intrinsic scatter of
the $M$--$T$ relation in a more subtle way that just accounted for by
cluster-to-cluster variations of the $\beta$--model fitting parameters
and of the effective polytropic index. The intrinsic scatter of the
$M$--$T$ relation is a further piece of information entering into the
determination of $\sigma_8$: since the theoretical mass function needs
to be convolved with this scatter, the high--temperature tail of the
model XTF grows with the scatter, at a fixed $\sigma_8$
(e.g. Pierpaoli et al. 2003). As a consequence, an overestimate of the
scatter turns into an underestimate of $\sigma_8$. We find that
$\sigma_8$ decreases by about 5 per cent by assuming 30 per cent,
instead of 16 per cent, intrinsic scatter. Although this effect only
partially compensates for the underestimate of the amplitude of the
$M$--$T$ relation, it highlights once again how precision measurements
of cosmological parameters require having under control a number of
systematic effects entering in the conversion between cluster mass and
observable quantities.
 
As a word of caution, we remind that the mismatch between $T_{\rm ew}$ and
$T_{\rm sl}$ is introduced by the complexity of the cluster thermal structure
and, therefore, so does the estimate of any biasing in the mass
estimate. Furthermore, in order to properly quantify the amount of a possible
bias in the observational determination of cluster masses it is required to
reproduce as closely as possible the observational procedure and the scale
range where the value of $\beta_{\rm fit}$ is fitted (see the discussion in
B04). We desert to a forthcoming paper (Rasia et al., in preparation) a
detailed analysis of the observational systematics in the determination of
cluster masses from X--ray data.  In this respect, it will be also important
to compare results on $\sigma_8$ from the XTF with those from other methods,
also based on galaxy clusters, but not relying on the choice of the $M$--$T$
relation. An example is provided by the approach based on the cluster baryon
mass function (Voevodkin \& Vikhlinin 2004), which provides $\sigma_8\simeq
0.7$.
 
\begin{figure}[t]  
\begin{center}  
\includegraphics[height=0.3\textheight,width=0.4\textwidth]{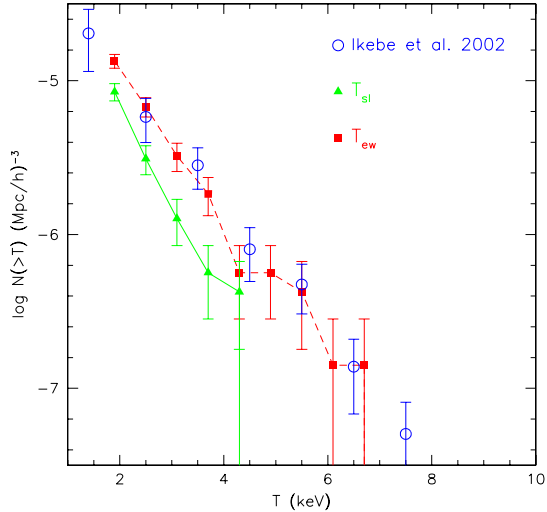}  
\caption{The cumulative temperature function  $N(>T)$, 
from Set\#1. Filled squares and dashed line refer to
the results for the emission-weighted temperature $T_{\rm ew}$, while
filled triangles and solid line represent the spectroscopic-like
temperature $T_{\rm sl}$. Error bars correspond to Poissonian
uncertainties. For reference, open circles show the local temperature
function measured by Ikebe et al. (2002) and adapted for the
considered cosmological model.  }
\label{fig:cumul}  
\end{center} 
\end{figure}

In Figure \ref{fig:cumul} we show the effect that using $T_{\rm sl}$,
instead of $T_{\rm ew}$, has on the estimate of the XTF. We compare
here the cumulative temperature function $N(>T)$ from the simulation
(Set\#1 only) to the data by Ikebe et al. (2002).  Using $T_{\rm ew}$
reproduces the observed XTF quite well (see also B04): this would
support $\sigma_8=0.8$, the value assumed in the simulation. However,
when $T_{\rm sl}$ is used to better account for the observational
procedure of temperature measurement, the simulation XTF is
significantly affected, especially in the high-temperature tail. This
result shows that $\sigma_8=0.8$ is excluded at 3$\sigma$ level, while
favoring a larger power spectrum normalization, at least for the ICM
physics introduced in our simulation.  For instance, at 2 keV the
simulated cumulative XTF is lower than the observed one by about 50-60
per cent. The value of $M_{\rm vir}$ corresponding to this temperature
in our simulation is about $1.5\times 10^{14}
h^{-1}M_\odot$. Computing the mass function by Sheth \& Tormen (1999)
for our cosmology at this mass, we find that $\sigma_8$ needs to be
increased to about 0.9 to increase the number density of clusters above this
mass limit by the required 50--60 per cent.

\section{Conclusions}\label{sec4} 
 
As demonstrated by Mazzotta et al. (2004; see also Mathiesen \& Evrard
2001), the emission-weighted temperature, $T_{\rm ew}$, usually
assumed to analyze the results of hydrodynamical simulations, fails in
reproducing the ``observational'' temperature obtained from the
spectral fit.  As an alternative, M04 proposed and tested a new
formula, named spectroscopic-like temperature, $T_{\rm sl}$, which
gives a much better approximation to the spectroscopic temperature,
with differences of a few per cent for $T>2$--3 keV.
 
In this Letter we have studied the consequences, for the
mass--temperature ($M$--$T$) relation and for the estimate of
$\sigma_8$ from the X--ray temperature function (XTF), of adopting
$T_{\rm sl}$, instead of $T_{\rm ew}$. We have analyzed a
set of hydrodynamical simulations of galaxy clusters, which include a
realistic treatment of the gas physics.  Our main results can be
summarized as follows.
\newline 
{\bf (a)} We find in our simulations that $T_{\rm sl}$ is smaller than
$T_{\rm ew}$ by a factor 0.7--0.8. A linear fit approximating the
relation between the two different temperature estimators is given by
Eq.~\ref{eq:TvsT}.
\newline 
{\bf (b)} Using $T_{\rm sl}$, instead of $T_{\rm ew}$, increases the
normalization of the $M$--$T$ relation making it higher than the
observed one by about 50 per cent.  By assuming hydrostatic
equilibrium for a gas density distribution described by a
$\beta$--model with a polytropic equation of state, masses are
underestimated on average by $\sim 40$ per cent, therefore
substantially reducing the discrepancy with observational data; the
scatter in the $M$--$T$ relation is also underestimated by about a
factor two.
\newline 
{\bf (c)} A bias in the $M$--$T$ relation propagates into a bias in
the power spectrum normalization, $\sigma_8$, from the XTF. If such a
bias is as large as that found in the simulations, the values
of $\sigma_8$ obtained by combining the local XTF and the observed
$M$--$T$ relation are underestimated by about 15 per cent.
\newline 
{\bf (d)} The XTF from the simulation is significantly lower when
using $T_{\rm sl}$ instead of $T_{\rm ew}$. A comparison with the
observed XTF (Ikebe et al. 2001) indicates that for the
``concordance'' $\Lambda$CDM model $\sigma_8$ needs 
to be increased from $\simeq 0.8$ to $\simeq 0.9$.

In general, our conclusions go in the direction of alleviating a
possible tension between the power--spectrum normalization obtained
from the number density of galaxy clusters and that arising from the
first--year WMAP CMB anisotropies (e.g. Bennett et al. 2003) and SDSS
galaxy power spectrum (e.g. Tegmark et al. 2004). Our results rely on:
{\em i)} the reliability of our simulations to correctly describe the
thermal complexity of the ICM; {\em ii)} the capability of reproducing
and understanding any possible source of bias in the estimates of
cluster masses. There is little doubt that these two issues need to be
addressed in detail in order to fully exploit the role that
hydrodynamical simulations play in the calibration of galaxy clusters
as precision tools for cosmology.
 
 
\acknowledgements{This work is supported CXC grants GO3-4163X and by 
European contract MERG-CT-2004-510143. Simulations were performed on
the CINECA IBM-SP4 supercomputing facility (INAF--CINECA grants). This
work is also supported by the the PD51 INFN grant. KD acknowledges
support by a Marie Curie fellowship of the European Community program
``Human Potential'' under contract number MCFI-2001-01221. We thank
S. Ettori, V. Springel, L. Tornatore and P. Tozzi for useful
discussions.  }

\end{document}